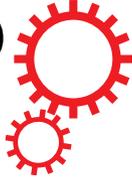

# Interface control by homoepitaxial growth in pulsed laser deposited iron chalcogenide thin films

Sebastian Molatta[1,2,3], Silvia Haindl[3,4], Sascha Trommler[2,3], Michael Schulze[2,3], Sabine Wurmehl[2,3] & Ruben Hühne[3]

Thin film growth of iron chalcogenides by pulsed laser deposition (PLD) is still a delicate issue in terms of simultaneous control of stoichiometry, texture, substrate/film interface properties, and superconducting properties. The high volatility of the constituents sharply limits optimal deposition temperatures to a narrow window and mainly challenges reproducibility for vacuum based methods. In this work we demonstrate the beneficial introduction of a semiconducting $FeSe_{1-x}Te_x$ seed layer for subsequent homoepitaxial growth of superconducting $FeSe_{1-x}Te_x$ thin film on MgO substrates. MgO is one of the most favorable substrates used in superconducting thin film applications, but the controlled growth of iron chalcogenide thin films on MgO has not yet been optimized and is the least understood. The large mismatch between the lattice constants of MgO and $FeSe_{1-x}Te_x$ of about 11% results in thin films with a mixed texture, that prevents further accurate investigations of a correlation between structural and electrical properties of $FeSe_{1-x}Te_x$. Here we present an effective way to significantly improve epitaxial growth of superconducting $FeSe_{1-x}Te_x$ thin films with reproducible high critical temperatures ($\geq 17$ K) at reduced deposition temperatures (200 °C–320 °C) on MgO using PLD. This offers a broad scope of various applications.

The superconducting iron-chalcogenide $FeSe_{1-x}Te_x$ has been intensively studied since 2008[1] and is rated as a candidate for high-field applications because of its high critical current densities up to magnetic fields of 30 T[2]. A high sensitivity of the superconducting transition temperature, $T_c$, to strain was reported as well[3,4]. A tunable $T_c$, for example, may be of advantage in future applications. Among the class of Fe-based superconductors the binary FeSe, FeTe, and the ternary $FeSe_{1-x}Te_x$ are structurally the most simplest compounds. They consist of stacked layers of $Fe_2X_2$ (X = chalcogen) where atoms are tetrahedrally coordinated (inverse PbO structure). Thus, it is widely believed as an easy-to-grow and easy-to-control system, however, thin film fabrication as well as the final processing of conductors still suffer serious difficulties[5]. The control of the stoichiometry, for example, is very challenging because of the high volatility of selenium and tellurium. A stoichiometric transfer at high deposition temperatures, $T_D$, is not guaranteed even using pulsed laser deposition (PLD). In order to counterbalance the volatility, recently $FeSe_{1-x}Te_x$ film deposition on $CaF_2$ at reduced substrate temperatures was investigated[6].

Several previous studies were devoted to a discussion of the correlation between structural and superconducting properties[7–18]. The first comparative study of $FeSe_{1-x}Te_x$ film growth on different substrates by Hanawa *et al.* proposed MgO, $CaF_2$ and $LaAlO_3$ as appropriate because no oxide interlayer formation was observed[12]. Despite the clean interface between $FeSe_{1-x}Te_x$ film and MgO substrate, the superconducting properties, especially $T_c$, is lowest when compared to $FeSe_{1-x}Te_x$ film deposition on other substrates. The highest $T_c$ for thin films of FeSe$_{1-x}$Te$_x$ grown by PLD reported so far reach 21 K on LaAlO$_3$[13], 23 K on CaF$_2$[14].

[1]Dresden High Magnetic Field Laboratory (HLD-EMFL), Helmholtz-Zentrum Dresden-Rossendorf, D-01314 Dresden, Germany. [2]Dresden University of Technology, Department of Physics, D-01062 Dresden, Germany. [3]IFW Dresden, P.O. Box 270116, 01171 Dresden, Germany. [4]Physikalisches Institut, Universitat Tubingen, Auf der Morgenstelle 14, D-72076 Tubingen, Germany. Correspondence and requests for materials should be addressed to S.M. (email: s.molatta@hzdr.de)

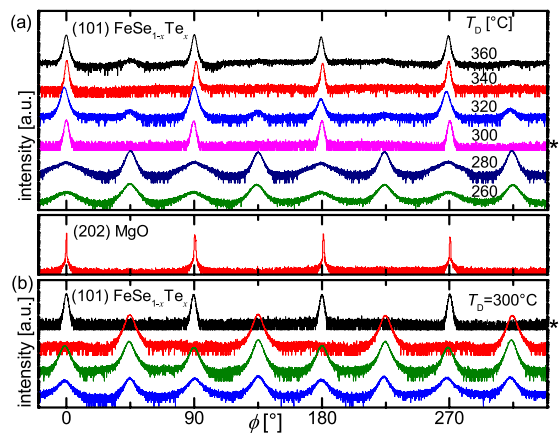

**Figure 1. *In-plane* texture of FeSe$_{1-x}$Te$_x$ thin films on MgO(001).** XRD $\phi$-scans of the FeSe$_{1-x}$Te$_x$ (101) reflection ($2\theta = 28.1°$, $\Psi = 57.0°$) (**a**) for different deposition temperatures. For comparison, $\phi$-scan of the MgO (202) reflection. (**b**) For various films grown at the same temperature, $T_D = 300°C$. Intensities are plotted logarithmically. The * indicates the same thin film.

However, the substrate/film interface lacks seriously homogeneity in stoichiometry and texture, especially for CaF$_2$ substrates, and thus challenges Fe chalcogenide and Fe pnictide thin film deposition for electronic applications. Attempts to better control nucleation exist for BaFe$_{2-x}$Co$_x$As$_2$ films such as the introduction of a SrTiO$_3$- ("template engineering")[19] or Fe-buffer[20]. In the case of iron chalcogenides a satisfactory solution is missing. Thersleff suggested the implementation of an iron buffer layer might also work for other Fe based superconductors which was shown by Iida *et al.*[21]. However, this iron buffer layer, a conductive shunt, displays ferromagnetic properties that have to be taken into account for interpretation of experimental results as well as for possible device operation. A beneficial diffusion barrier was introduced by Ichinose *et al.*[22] for the FeSe$_{1-x}$Te$_x$ thin film growth on CaF$_2$ substrates. In the case of the deposition on MgO substrates, the seed layer is clearly not a diffusion barrier, but decreases the lattice constant mismatch, enables deposition at lower temperatures and, therefore, results in better controllable growth of superconducting FeSe$_{1-x}$Te$_x$ thin films.

MgO is one of the most favorable substrates for a broad range of applications, has certain advantages compared to other substrates and is frequently used in thin film applications: (a) it is a cheap material with high thermal stability and chemical compatibility; (b) it has furthermore a low dielectric constant and low dielectric losses favorable for applications based on high frequencies. Finally, MgO is easily deposited biaxially textured on technologically useful templates such as Hastelloy tapes via the ion-beam assisted deposition (IBAD) process as used for high current carrying conductors. However, as already anticipated above, the direct deposition of FeSe$_{1-x}$Te$_x$ on MgO is characterized by diverging results and insufficient reproducibility.

In this work the beneficial implementation of a semiconducting FeSe$_{1-x}$Te$_x$ seed layer for subsequent homoepitaxial growth of a superconducting FeSe$_{1-x}$Te$_x$ film is presented. As we will demonstrate, this novel approach allows to improve control and reproducibility of structural and superconducting properties of FeSe$_{1-x}$Te$_x$ thin films. Unlike the iron buffer layer, the FeSe$_{1-x}$Te$_x$ seed layer shows no metallic, but semiconducting behavior at low temperatures without detrimental ferromagnetism. Furthermore, it may be advantageous for conductor fabrication on IBAD-MgO or other applications.

## Results

**Films grown directly on MgO(100).** We first analyze FeSe$_{1-x}$Te$_x$ film growth on MgO(001) substrates. In order to find optimal deposition parameters, we started with the conditions reported as optimum by Bellingeri *et al.*[13,23]. In contrast to this reference we did not find superconductivity in films grown above a deposition temperature of 380°C. In addition various *out-of-plane* orientations different from (00*l*) were observed for $T_D > 430°C$ (see supplement S I). Below this temperature down to 280°C the films are *c*-axis textured. However, the *in-plane* orientation of the FeSe$_{1-x}$Te$_x$ [Fig. 1(a)] commonly show two distinguished epitaxial relationships, as X-ray diffraction (XRD) confirms. In particular for $T_D \leq 300°C$: grains grown *cube-on-cube*, i.e. (001)[100]FeSe$_{1-x}$Te$_x$ ∥(001)[100]MgO, and grains with 45° *in-plane* rotated orientation (001)[110]FeSe$_{1-x}$Te$_x$ ∥(001)[100]MgO are found. Although two films grown at 340°C and at 300°C show only a single epitaxial relationship, a more detailed investigation on the reproducibility performed at $T_D = 300°C$ demonstrates a lack of controllability and a significant variance in the results [Fig. 1(b)]. In addition, for films grown below 300 °C, the mosaicity becomes significantly larger (see supplement S II).

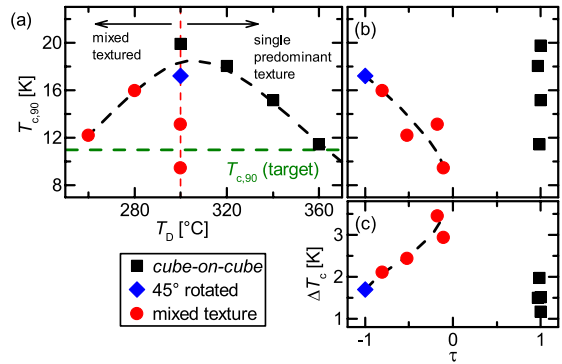

**Figure 2. Dependence of the critical temperature on $T_D$ and texture.** (**a**) $T_{c,90}$ vs. $T_D$ of thin films deposited directly on MgO , the horizontal dashed line indicates $T_{c,90}$ of the ablated target. (**b**) $T_{c,90}$ vs. texture coefficient $\tau$. (**c**) $\Delta T_c$ vs. $\tau$. The black dashed lines are guides to the eye.

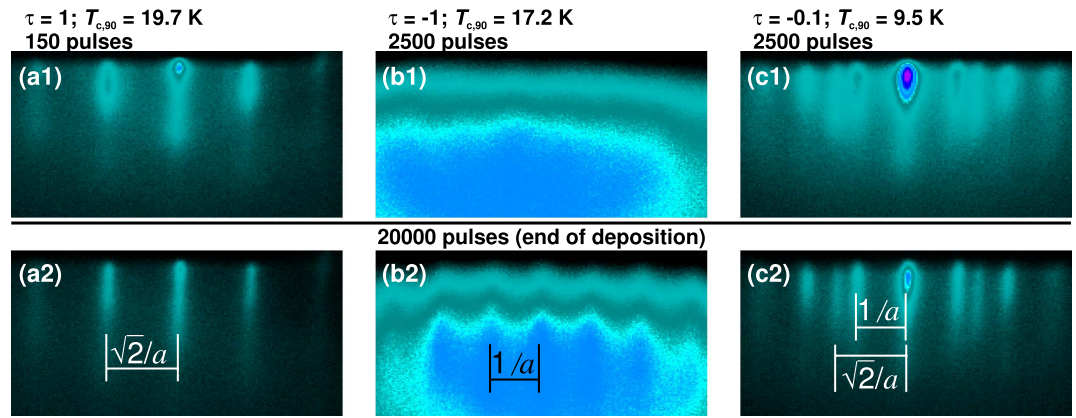

**Figure 3. RHEED-images of thin films deposited at 300 °C.** (a1),(b1),(c1) at the beginning of the deposition; (a2),(b2),(c2) at the end of deposition. The electron beam impinges the sample along the (110) substrate-plane in an angle of less than 3°.

In order to compare the texture of different thin film samples and quantify the degree of texture admixture we introduce a texture coefficient

$$\tau = \frac{I_{[100]} - I_{[110]}}{I_{[100]} + I_{[110]}} \tag{1}$$

with $I_{[100]}$ and $I_{[110]}$ the integral intensity of the *cube-on-cube* grown and the 45° *in-plane* rotated grown texture component respectively. Based on this texture coefficient we classify the samples as follows. For $|\tau| > 0.95$ the sample is nearly or fully *in-plane* textured, for $|\tau| < 0.95$ the thin film has a mixed texture. For $\tau > 0$ the *cube-on-cube* grown texture component is predominant and for $\tau < 0$ the 45° rotated component prevailed.

For $T_D > 320$ °C the structural and electrical properties of the grown thin films are in good accordance to the results of Huang *et al.*[11]. As is evident from our investigation the superconducting transition temperature increases with decreasing $T_D$ and has a maximum of $T_{c,90} = 19.7$ K at $T_D = 300$ °C [Fig. 2(a)]. However in this regime of $T_D$ the texture is not well controllable obstructing the reproducibility [Fig. 1(b)].

By comparing structural properties and $T_{c,90}$ (supplement S III) we can classify different regimes: FeSe$_{1-x}$Te$_x$ films with $T_{c,90}$ comparable or above the $T_{c,90}$ of the target can be grown between $T_D = 260$ °C and 360 °C [Fig. 2(a)]. However, epitaxial film growth is only favored above 300 °C. At 300 °C and below, a second grain orientation (45° -rotated) starts to significantly compete with the *cube-on-cube* texture. This competition is already observable from grain nucleation, as is confirmed by reflection high energy electron diffraction (RHEED) [Fig. 3]. For films with $0.95 < |\tau| < 0$ (mixed texture) $T_{c,90}$ increases with increasing $|\tau|$ [Fig. 2(b)]. Correspondingly, the superconducting transition width, $\Delta T_c = T_{c,90} - T_{c,10}$, increases with increasing texture admixture (decreasing $|\tau|$) as shown in Fig. 2(c). Notable is that in mixed textured samples the 45° *in-plane* rotated texture component always is dominant.

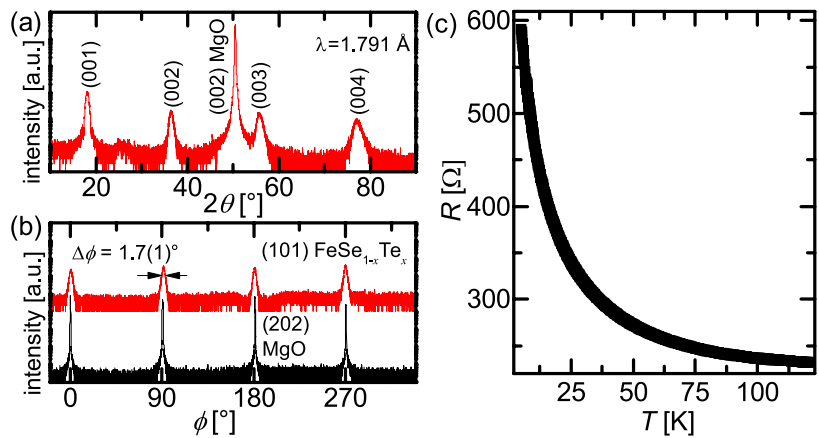

**Figure 4.** FeSe$_{1-x}$Te$_x$ seed layer characterization: **(a)** XRD $\theta-2\theta$ -scan of a seed layer with pure *c*-axis orientation out-of-plane. **(b)** XRD $\phi$-scan of a seed layer with the epitaxial relation (001)[100]FeSe$_{1-x}$Te$_x$ ∥(001)[100]MgO. Intensities are plotted logarithmically. **(c)** $R$ vs. $T$ of the 20 nm FeSe$_{1-x}$Te$_x$ seed layer grown on MgO(001).

From RHEED imaging [Fig. 3] information on the structure of the thin films is obtained during film growth. We compare three films made at the same deposition conditions ($T_D = 300$ °C).

The RHEED images of the thin film with $\tau = 1$ show clearly visible reflections (streaks) directly from the beginning [Fig. 3(a1)] until the end [Fig. 3(a2)] of the deposition. This indicates a very fast nucleation and formation of the texture as well as a very smooth surface. From the spacing of the streaks ($\sqrt{2}/a$) the *in-plane* orientation can be determined as *cube-on-cube*.

In Fig. 3(b1),(b2) RHEED images of the film with $\tau = -1$ are shown. In contrast to the *cube-on-cube* grown thin film, even after 2500 pulses the RHEED pattern is very blurred and the streaks are hard to determine. This indicates that the nucleation during the growth of this sample took much longer time than at the deposition of the thin film with $\tau = 1$. After 20000 pulses the streaks are more pronounced but still very blurry reflecting the high mosaicity and the broad peaks of the $\phi$-scans. The distance of the streaks ($1/a$) confirms a 45° rotated orientation of the FeSe$_{1-x}$Te$_x$ thin film with respect to the substrate orientation. The absence of additional reflexes is again an evidence for a smooth surface.

The RHEED images of the highly mixed textured sample ($\tau = -0.1$) shows streaks for both orientations, *cube-on-cube* grown ($\sqrt{2}/a$) and 45° rotated ($1/a$), already at the beginning of the deposition [2500 pulses, Fig. 3(c1)]. These streaks are quite blurry but still well pronounced and become sharper with the ongoing deposition. At the end of the deposition [Fig. 3(c2)] the reflections of both orientations are very sharp and very pronounced. The slightly higher intensity of the reflections of the 45° rotated grains compared to the intensity of the streaks of the *cube-on-cube* grown grains of the sample reders the ratio of these two fractions ($\tau \approx -0.1$). In the RHEED pattern of this thin film there are also no additional reflexes observable.

**Homoepitaxial growth of thin FeSe$_{1-x}$Te$_x$ films using a seed layer.** In order to control epitaxial growth at low $T_D$ and simultaneously ensure high $T_{c,90}$ we propose a novel method that is based on homoepitaxial growth of FeSe$_{1-x}$Te$_x$. Therefore, a thin FeSe$_{1-x}$Te$_x$ seed layer is deposited at 400 °C on MgO(001). This epitaxially grown (*cube-on-cube*) layer is semiconducting [Fig. 4]. The normal state resistance of the seed layer is ≈200 Ω at 100 K [Fig. 4(c)]. The normal state resistance of a superconducting FeSe$_{1-x}$Te$_x$ layer is ≈10 to 20 Ω at 100 K regardless of whether it is deposited directly on a MgO substrate or on a seed layer. The ratio of $R_{sc\ layer}/R_{seed}$ is ≈1/20 at 100 K and decreases with decreasing temperature. Therefore, the seed layer has negligible detrimental effects on further transport investigations.

Subsequent to the growth of the seed layer, the temperature of the substrate is reduced and another FeSe$_{1-x}$Te$_x$ layer is deposited. As confirmed by XRD measurements the FeSe$_{1-x}$Te$_x$ layers are grown with a single texture component with the epitaxial relation: (001)[100]FeSe$_{1-x}$Te$_x^{sc\ layer}$ ∥(001)[100]FeSe$_{1-x}$Te$_x^{seed}$ ∥(001)[100]MgO [Fig. 5]. For the full width at half maximum (FWHM, $\Delta\phi$) of the peaks of the *in-plane* orientation, a strong increase is observable with decreasing $T_D$ in the sample series without seed layer. In contrast, the $\Delta\phi$ of the seeded films is more or less constant until a $T_D$ of 200 °C (see supplement S II).

Seeded films furthermore have a smoother surface as atomic force microscope (AFM) images support (see supplement S IV). The root-mean-square (rms) roughness was evaluated with the WSXM software package[24]. The film without seed layer ($T_D = 300\,°\text{C}$, $\tau = 1$) has a rough surface (rms: 0.88 nm) compared to a very smooth profile of the film with seed layer ($T_D = 240\,°\text{C}$, rms: 0.38 nm).

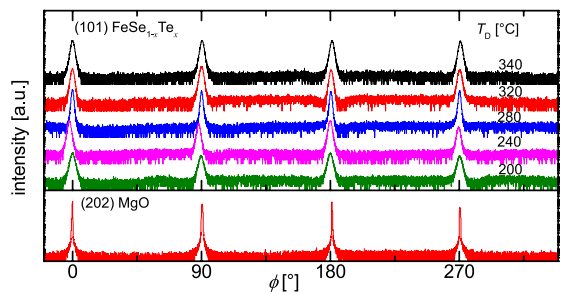

**Figure 5. XRD $\phi$-scans of films homoepitaxially grown on a seed layer in a temperature series (200 °C–340 °C).**

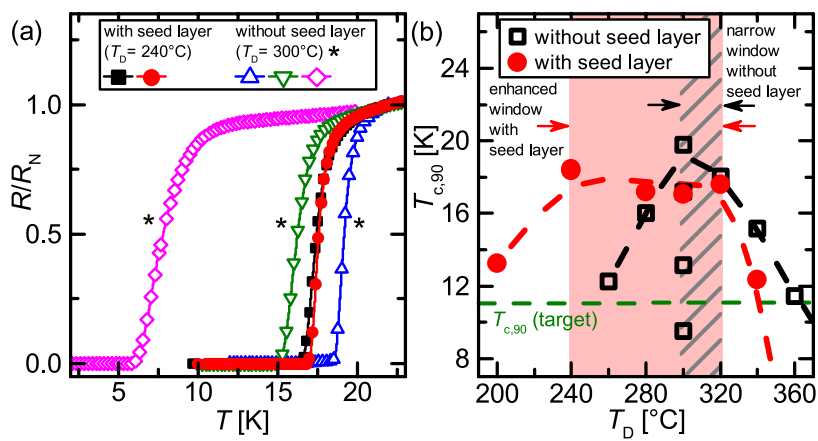

**Figure 6.** (**a**) Temperature dependence of the normalized resistance ($R_N = R(22\,\text{K})$) for thin films without ($T_D = 300\,°\text{C}$) and with ($T_D = 240\,°\text{C}$) seed layer. The $^*$ marks the samples without seed layer. (**b**) Comparison of $T_{c,90}$ of the samples with and without seed layer in dependence of $T_D$. The gray shaded area indicates the deposition temperature window for high $T_{c,90}$ thin films without seed layer, the light red area indicates the $T_D$ window for high $T_{c,90}$ samples with seed layer. The horizontal dashed line indicates $T_{c,90}$ of the target, the black and red dashed lines are guides to the eye.

Electrical resistance measurements [Fig. 6(a)] for different $T_D$ where the highest $T_{c,90}$ was found for thin films with ($T_D = 240\,°\text{C}$) and without ($T_D = 300\,°\text{C}$) seed layer demonstrate the improved reproducibility using the seed layer. The curves of the samples without seed layer show a large spread in $T_{c,90}$ as well as of the superconducting transition width. In contrast to that, results for seeded films were fully reproducible [black and red solid symbols in fig. 6 (a)].

Improved control of $T_{c,90}$ is gained in seeded films for low deposition temperatures. The optimal deposition temperature window broadens (240 °C–320 °C) for films with high $T_{c,90}$ ($\geq 17\,\text{K}$) as can be seen in Fig. 6(b). Even though the films grown at $T_D > 320\,°\text{C}$ show only a single epitaxial relation to the substrate with sharp reflections in the XRD scans [Fig. 5, supplement Fig. S 2], indicating a high crystalline quality, the thin films show a low $T_{c,90}$ ($<17\,\text{K}$) similar to the high-$T_D$ samples without seed layer, assuming an additional mechanism partially suppressing superconductivity in these films.

## Conclusion

To summarize, we have overcome some challenges reported for the thin film growth of iron chalcogenides using PLD. We have demonstrated that homoepitaxial FeSe$_{1-x}$Te$_x$ thin film growth (i.e. the use of a FeSe$_{1-x}$Te$_x$ seed) with the above described procedure is a very powerful method for optimization with respect to reproducibility.

High deposition temperatures result in high mobility of the ions on the substrate surface. The resulting deposited layer thus is epitaxially grown but does not show superconductivity. When used as a seed layer, a subsequent deposition of the same target material at lower temperatures results in a homoepitaxial growth of superconducting FeSe$_{1-x}$Te$_x$ films. Due to the smaller lattice mismatch between FeSe$_{1-x}$Te$_x$ and the FeSe$_{1-x}$Te$_x$ seed compared to MgO and, possibly, to a much better chemical compatibility, epitaxial growth is ensured at lower temperatures.

The method presented in this article opens a significantly broader window of deposition temperatures for high-$T_{c,90}$ films and ensures good epitaxial growth. We also found, that in a wide interval (240 °C–320 °C) the deposition temperature is not crucial for the superconducting properties of the thin film. Furthermore it grants access to MgO which is one of the most favorable single crystal substrates for thin film fabrication. Because of the similar structur within the *ab-plane*, we predict that this semiconducting seed layer will also be adoptable for other iron based superconductors more difficult to grow using PLD. This technique is easy to implement in the film growth process and may be of advantage for a vast variety of applications.

## Methods

The analyzed thin film samples in this report were prepared by PLD using a Lambda Physics LPX 305 KrF-excimer laser ($\lambda = 248$ nm, $\varepsilon = 3 - 5$ J/cm$^2$) in an ultra-high-vacuum-chamber with a base pressure of $10^{-9}$ mbar. The target - substrate distance is about 40 mm. The used targets were grown by a modified Bridgman-process with a nominal stoichiometry of Fe:Se:Te = 2:1:1 resulting in a bulk $T_{c,90}$ of 11 K[25,26].

For the direct (unseeded) preparation of the samples the target material was deposited on the substrate at deposition temperatures, $T_D$, from 260 °C to 530 °C at a laser repetition rate of 7 Hz. For the seeded FeSe$_{1-x}$Te$_x$ films, a FeSe$_{1-x}$Te$_x$ layer with a thickness of about 20 nm was deposited at 400 °C with a laser repetition rate of 10 Hz. The as-grown film was cooled to the deposition temperature in the range of 200 °C–340 °C for a subsequent homoepitaxial FeSe$_{1-x}$Te$_x$ layer, deposited at a laser repetition rate of 7 Hz.

The superconducting films have a thickness of about 120 nm ± 30 nm confirmed by film cross-sections cut by a focused ion beam (Ga ions) and imaged by scanning electron microscopy. The film growth was monitored *in-situ* by RHEED evaluated with the KSA400 software.

XRD $\theta-2\theta$-scans were measured in a Philips X'Pert with Cu-K$\alpha$ or Co-K$\alpha$-radiation. The $\phi$-scans for the (101) peak of FeSe$_{1-x}$Te$_x$ were measured in a Philips X'Pert PW3040 equipped with a four circle goniometer with Cu-K$\alpha$-radiation in Bragg-Brentano-geometry.

$R(T)$ measurements were made in a Quantum Design Physical Properties Measurement System (PPMS). $T_{c,90}$ describes the temperature where the resistance reached 90% of the normal state resistance above transition.

Surface morphology was examined with an Digital Instruments AFM Dimension 3100 with a Nanoscope IIIA controller in tapping mode.

## Acknowledgements

S.M. thanks L. Schultz for supervision, J. Hänisch, K. Iida, F. Kurth, S. Oswald and S. Kaschube for fruitful discussions, M. Kühnel and J. Scheiter for technical support. S.M. acknowledges financial funding from DFG-GRK 1621, S.H. from DFG (project HA5934/3-2), and SW from DFG in project WU595/3-1 (Emmy-Noether project) and the SPP (BU887/15-1) and by Bundesministerium für Bildung und Forschung (BMBF) in the ERA.Net RUS program (project #245 FeSuCo).

## Author Contributions

S.M. and S.T. have grown the films and conducted the experiments, S.M., S.H., S.T., and R.H. analyzed the results, S.M. and S.H. have written the manuscript, M.S. and S.W. provided the target material, and R.H., S.T., and S.M. designed the experiment. All authors have discussed the results.

## Additional Information

**Supplementary information** accompanies this paper at http://www.nature.com/srep

**Competing financial interests:** The authors declare no competing financial interests.

**How to cite this article**: Molatta, S. *et al.* Interface control by homoepitaxial growth in pulsed laser deposited iron chalcogenide thin films. *Sci. Rep.* **5**, 16334; doi: 10.1038/srep16334 (2015).

# Supplementary Information

# Interface control by homoepitaxial growth in pulsed laser deposited iron chalcogenide thin films

Sebastian Molatta[1,2,3,*], Silvia Haind[3,4], Sascha Trommler[2,3], Michael Schulze[2,3], Sabine Wurmehl[2,3], Ruben Hühne[3]

[1] Dresden High Magnetic Field Laboratory (HLD-EMFL), Helmholtz-Zentrum Dresden-Rossendorf, D-01314 Dresden, Germany

[2] Dresden University of Technology, Department of Physics, D-01062 Dresden, Germany

[3] Leibniz-Institute for Solid State and Materials Research (IFW) Dresden, Helmholtzstraße 20, D-01069 Dresden, Germany

[4] Physikalisches Institut, Universität Tübingen, Auf der Morgenstelle 14, D-72076 Tübingen, Germany

[*] corresponding author's e-mail address: s.molatta@hzdr.de

## S1 XRD of unseeded film at high deposition temperature

In Fig. S1 the XRD $\theta$–$2\theta$-scans of films grown directly on MgO at $T_D$'s of 530 °C and 300 °C are shown. Besides $c$-axis grown components of the film with $T_D = 530$ °C ((00$l$)-reflections) there are additional *out-of-plane* orientations observable. Peaks of the ($h0l$) as well as the ($hkl$) orientation are found [Fig. S1(a)]. With decreasing $T_D$ the intensity of the peaks for orientations different from (00$l$) decreases rapidly. The XRD $\theta$–$2\theta$-scan of the film with $T_D = 300$ °C shows exclusively $c$-axis orientation *out-of-plane* [Fig. S1(b)].

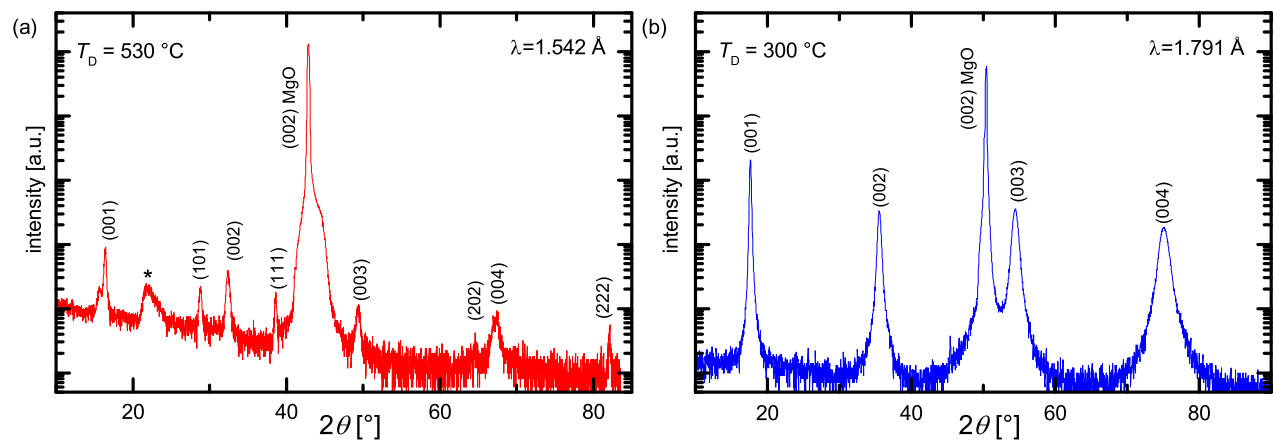

**Figure S1.** XRD $\theta$–$2\theta$-scan of FeSe$_{1-x}$Te$_x$ thin films grown on MgO at (a) $T_D = 530$ °C and (b) $T_D = 300$ °C. The $\star$ indicates the $\lambda/2$-peak of MgO.

## S II  In-plane texture: Comparison between unseeded and seeded films on MgO

Evaluating the FWHM of the peaks of the *in-plane* orientation of the temperature series [Fig. S 2] shows a strong increase of $\Delta\phi$ for both texture components for the thin films without seed layer for decreasing $T_D$. In contrast to that, the $\Delta\phi$ values of thin films with seed layer stay more or less constant with decreasing deposition temperature.

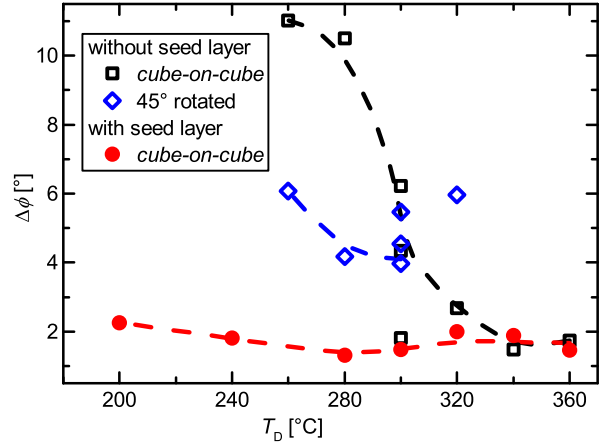

**Figure S 2.** $\Delta\phi$ of the texture components of the samples without and with seed layer in dependence of $T_D$.

## S III  Resistance measurements of films without seed layer

From the resistance measurements [Fig. S 3] $T_{c,90}$ and $\Delta T_c$ for the thin films without seed layer are obtained. $\Delta T_c$ is calculated from the temperature where the resistance reaches 90% of the normal state resistance minus the temperature where the resistance reaches 10% of the normal state resistance.

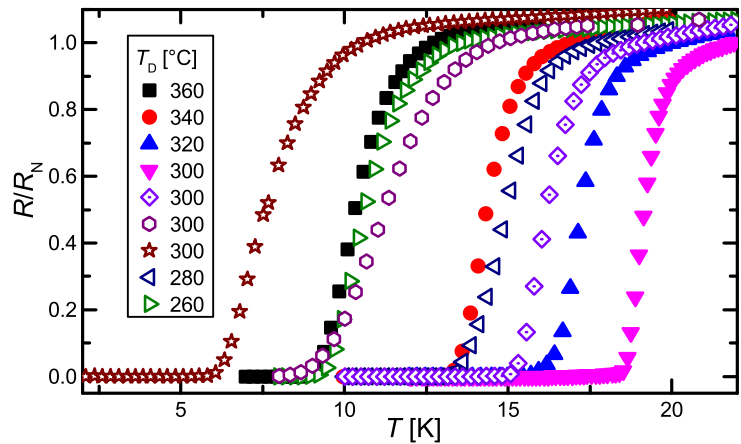

**Figure S 3.** Temperature dependence of the normalized resistance ($R_N = R(100\%)$) for thin films without seed layer.

## S IV  Comparison of the surface morphology

The surface morphology of a sample without seed layer [Fig. S 4 (a) $T_D = 300\,°C$, $\tau = 1$] and a sample with seed layer [Fig. S 4 (b) $T_D = 240\,°C$] was evaluated with AFM. The rms-values of the samples show an improvement of the surface roughness by a factor of 2.

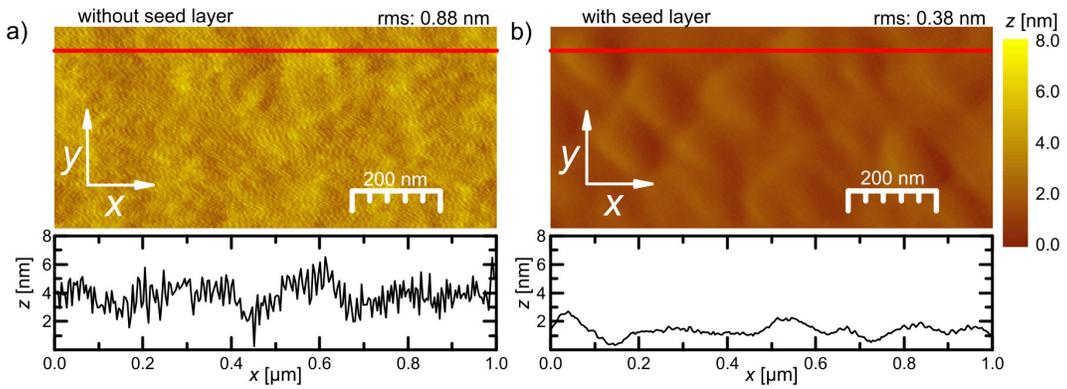

**Figure S 4.** AFM images of a thin film grown without seed layer [(a) $T_D = 300\,°C$, $\tau = 1$] and a film on the seed layer [(b) $T_D = 240\,°C$]. The red lines mark the position of the profiles below.